High energy resolution GeV gamma ray detector

Satoko Osone

Funabashi, Chiba, 273-0865, JAPAN

osone@icrr.u-tokyo.ac.jp

Abstract

We design a GeV gamma ray detector based on an electron-positron pair measurement by using the BESS-polar magnet and study it with the GEANT4 simulation code. We consider an additional counter to select an electron-positron pair for which the energy loss by bremsstrahlung is less than 100 MeV against AMS. We obtain a significantly low detector efficiency of 3%. However, an energy resolution better than 1% at 10–100 GeV is achieved by employing a parameter that enables us to obtain a line sensitivity that is almost indetical to that of the GLAST mission. The good energy resolution makes it possible to obtain the three-dimensional map of the neutralino dark matter annihilation line.

1. Introduction

The science of GeV gamma rays, blazars and diffuse gamma ray backgrounds has been studied with the EGRET on the *CGRO* satellite. The next-generation detector GLAST has a greater effective area, better energy resolution, better angular resolution as compared to the EGRET and it has a sensitivity up to 100 GeV. The ground detector of the air Cherenkov method MAGIC has a sensitivity above 50 GeV. The energy spectra in a wide energy range can be observed with a combination of a space observation and a ground observation. Such energy spectra are needed to discriminate between electron acceleration and proton acceleration for both blazars and supernova remnants. Next-generation detectors such as MAGIC and GLAST will also study emission from gamma ray bursts at GeV-TeV energies and a neutralino dark matter annihilation line that is expected at 10–100 GeV.

Original detectors of GeV gamma rays in space are composed of a tracker for determining the arrival direction and damping gamma ray energy and a calorimeter for the measurement of energy. There is a new method to detect GeV gamma rays in space [1][2]. In this method, a gamma ray is converted into an electron-positron pair by a very thin convertor. Further, the tracks of the electron and positron are measured with a Si strip in a magnet in order to obtain the momentum of the pair and the energy of gamma ray is estimated by applying the principle of conservation of energy. We call this method as a track mode. AMS has two modes—a calorimeter mode and a track mode for gamma ray observation [3]. AMS uses TRD ($0.25 X_0$) as a convertor. We design a GeV gamma ray detector by using a track mode by employing the BESS-polar magnet and study it with the GEANT4 simulation code. We find that the energy loss of an electron and a positron by bremsstrahlung in the convertor is important. We consider an additional counter against AMS in order to select an electron-positron pair with low energy loss.

2. Description of detector

Our detector is composed of an anti counter, a convertor, a tracker in a Magnet, and an additional counter, as shown in figure 1. We consider the use of the International Space Station.
We eliminate cosmic rays by using a plastic scintillator. When there is an energy deposit in the plastic scintillator, we regard it to be done to cosmic rays. We use BESS-polar magnet [4], that is 0.8 m in diameter and 1.4 m in length. This Magnet has been used with a magnetic field of 1 T, while we use it with a magnetic field of 2 T in order to obtain a better energy resolution. Magnet become $\sqrt{2}$ times as thick as BESS-polar with a magnetic field of 1 T, 4.84 mm ($0.085\ X_0$). The thickness of a cryostat is $0.056\ X_0$. The total quantity of material is $0.141\ X_0$. We consider the combination of a magnet and a cryostat for constructing the convertor. We prepare six layers of Si strips in a magnet. The size of a Si strip layer is $0.6 \times 1$ m. We have to eliminate a magnetic force in space. Therefore, we prepare two magnets because magnet of BESS-polar is solenoid. Therefore, the total geometrical area of our detector is $2 \times 0.6$ m$^2$ = 1.2 m$^2$.

For a track mode, multiple scattering and the energy loss of an electron and a positron by bremsstrahlung are important.

Initially, we study a multiple scattering if we can obtain a clear track in the magnet. The deflection angle by a magnetic force is given by $\phi_{mag} = LeB / pc = 300\ B(\text{T})L(\text{m}) / pc(\text{MeV})$. Here, $L$ is the traverse length; $B$, the magnetic field; $pc$, the momentum of the particle. The deflection angle due to multiple scattering is given as $\phi_{scatt} = Es \sqrt{l(\text{m})/X_0(\text{m})} / \sqrt{2}\ p\beta c$. Here, $Es$ = 21 MeV, $X_0$ is the radiation length, and $l$ is the thickness of material. The ratio of these two deflection angles is given as $\phi_{scatt}/\phi_{mag}$ = 0.05 $\sqrt{l(\text{m}) / X_0(\text{m})} / B(\text{T})\beta L(\text{m})$. The material comprises the magnet, cryostat (0.14 $X_0$) and six layers of Si strips (0.03 $X_0$). Here, the thickness of one Si strip layer is 500 μm. For a relativistic particle ($\beta$ = 1), $B$ = 2 T, $L$ = 0.8 m and $l$ = 0.17 $X_0$, the ratio of the deflection angles becomes 0.013. We simulate the deflection with a magnetic field and without a magnetic field by using the GEANT4 simulation code. The deflection without a magnetic field corresponds to multiple scattering. We check the deflection at the sixth Si layer. The deflection without a magnetic field is 0.12 cm, 0.026 cm, 30.4 μm, and 1.06 μm for 1-, 10-, 100-GeV, and 1-TeV electron, respectively. The deflection with a magnetic field is 22.1 cm, 2.02 cm, 2.02 mm, and 201 μm for 1-, 10-, 100-GeV, 1-TeV electron, respectively. The ratio becomes 0.005, 0.013, 0.015, and 0.005 for 1-, 10-, 100-GeV, and 1-TeV electron, respectively. We confirm the calculated result with the aid of the simulation. Therefore, the effect of multiple scattering on the track can be neglected.

Further, we study the energy loss of an electron and a positron by bremsstrahlung in the magnet and cryostat (0.14 $X_0$). We simulate the conversion of 30000 gamma rays of 100 GeV energy by using the magnet and cryostat. The statistics are almost the same when observations of Galactic diffuse gamma ray backgrounds for a geometrical area of 1 m² str are considered for a period of 1 yr. We show the constructed energy of an electron-positron pair in figure 2. We select the energy of an electron and a positron to be above 1 GeV because a lower-energy particle creates a double track in a Si strip and we fail a track reconstruction. There are a peak around 100 GeV and a long tail to 1 GeV in figure 2. Eight percent of the gamma rays form an electron-positron pair and 46% of electron-positron pairs lose an energy less than 100 MeV. We have to produce an energy response when we reconstruct the gamma ray energy spectra because an electron and a positron lose energy in the convertor. However, it is difficult to produce a precise energy response, particulary for low statistics gamma ray because the constructed energy distribution of an electron and a positron is discontinuous.

We consider an additional counter to detect the gamma rays from bremsstrahlung that have an energy above a particular energy. This counter is composed of an absorber and a tracker, as shown in figure 3. The absorber is the bottom of the magnet and cryostat (0.14 $X_0$) and an additional lead. The tracker is the sixth layer of the Si strip in the magnet and an additional Si. The additional Si and lead are prepared with several layers. When an electron or a positron passes a tracker, there is an energy deposit. When gamma rays from bremsstrahlung pass the sixth Si strip layer in the magnet, there is no energy deposit. By interacting with the lead, a gamma ray from bremsstrahlung makes an electron-positron pair and there is an energy deposit when they pass the next absorber. In order to discriminate between an electron-positron pair and a gamma ray from bremsstrahlung, we construct the three-dimensional image of a Si strip. We prepare 1 layer of Si and lead, input 100 gamma rays as bremsstrahlung with energies from 10 MeV to 100 MeV, and show a number of a hit in a Si strip in figure 4. We set three kinds of a thicknesses for the lead—1 mm (0.2 $X_0$), 5.5 mm (1 $X_0$), 1 cm (5 $X_0$). A hit by the low-energy gamma rays of bremsstrahlung is small because the cross section of a pair creation is small for low-energy gamma rays. When the thickness of the lead is large, the gamma rays lose total energy in a lead and there is no energy deposit in a Si strip. When the thickness of the lead is small, the number of hits is small because the probability of a pair creation is small. We prepare two kinds of counter, six layers of Si and lead of 5.5 mm, and five layers of Si and lead of 1 cm, input 100 gamma rays as bremsstrahlung with energies from 10 MeV to 1 GeV, and show a number of a hit in figure 5.  Niety-six percent of the 100-MeV gamma rays from bremsstrahlung make a hit with the six layers of a lead of 5.5 mm and Si. We cannot increase the number of layers infinitely because of limit on the weight and money. Therefore, we set six layers of Si strips and a lead of 5.5 mm as an additional counter and set the threshold energy as 100 MeV. We can obtain the energy of 10-GeV gamma rays with an accuracy of 1%. In addition to this counter, we planned to produce an energy response. This further improves the energy resolution.

3. Quantum efficiency of detector

We study the efficiency of the detector by using the GEANT4 simulation code.

Initially, we study the efficiency for a lead, magnet and cryostat that functions as a convertor. We input 1000 gamma rays with energies from 5 GeV to 1 TeV and simulate their conversion in the convertor; we represent the number of selected events as a function of the gamma ray energy in figure 6. We select an event satisfying the condition that the energy of bremsstrahlung gamma rays is less than 100 MeV and the number of charged particles is two (corresponding an electron-positron pair). We add the selection condition that the energy of an electron and a positron should be above 1 GeV because a lower-energy particle makes a double track in a Si strip and we fail a track reconstruction. We set eight thicknesses for the lead—0, 100 μm, 500 μm, 1 mm, 2 mm, 3 mm, 4 mm, and 5 mm. One radiation length for a lead is 5.5 mm. A thickness of zero indicates that the lead is of no use. If a convertor is thin, the probability of pair creation is low; however, the energy loss of pair is small. If a convertor is thick, the probability of pair creation is large; however, the energy loss of pair is large. We found that the lead does not cause the number of hits to increase and hence we do not need a lead. We obtain a quantum efficiency of 4 % for a magnet and a cryostat.

Further, we study the efficiency of a Si tracker for an electron and a positron. We input 100 particles with energies from 1 GeV to 1 TeV. We show the number of selected events as a function of particle energy in figure 7. We select an event satifying the condition that the energy of bremsstrahlung gamma rays is less than 100 MeV and the number of charged particles is 1, either as an electron or a positron. We find the the quantum efficiency in the Si tracker is 80% for an electron and a positron. Therefore, the total quantum efficiency of our detector for gamma rays is 4 × 0.8 × 0.8 = 3%.

4. Comparion with other detectors

4.1 Energy resolution

There are two kinds of limits that determine the energy resolution. One is the accuracy in the determination of an energy loss by using an additional counter, and the other is the track accuracy. We detect bremsstrahlung above 100 MeV with an additional counter. Therefore, the accuracy of the energy determination is 1% for 10 GeV and 0.1% for 100 GeV. The energy resolution by the track accuracy is defined as follows [5]. $\Delta P/P = \sigma(x)(m)\ p(GeV/c)\ \sqrt{720/(N+4)}\ / 0.3\ B(T)\ L(m)^2$. Here, $\sigma(x)$ is the precision of the position; $p$, the momentum of the particle; $N$, the number of trackers; $B$, the magnetic field; and $L$, the traverse length. We adopt $B = 2$ T, $\sigma(x) = 5$ μm, $N = 6$, and $L = 0.8$ m compared the results with those for $B = 0.8$ T, $\sigma(x) = 10$ μm, $N = 8$, and $L = 1$ m for the AMS. We show an energy dependent energy resolution in figure 8. Our detector has an energy resolution allowed by two limits. In the range 10–100 GeV, the energy resolution is superior to that of GLAST and AMS.

4.2 Line sensitivity

We show the effective area in figure 9 which is approximately 0.04 m$^2$; this is 1/25 times smaller than that of GLAST. The field of view is approximately 2 str. With a 3 yr. observation, the statistics of Crab is 12 photon at 1 GeV. A small effective area makes the observation of a point source impossible. The science with this detector is diffuse gamma ray background, bright GRB and neutralino dark matter annihilation line. With 3 yr. observation, the statistics of Galactic diffuse gamma ray background is 90000 photon and 9000 photon at 10 GeV and 100 GeV, respectively. With 3 yr. observation, the statistics of extragalactic diffuse gamma ray background is 9000 photon and 900 photon at 10 GeV and 100 GeV, respectively.

First, we discuss the feasibility of observing a neutralino dark matter annihilation line with this detector. Radaz calculated the flux of neutralino dark matter annihilation line[6]. Recent papers show that there is an enhancement of the statistics of the neutralino dark matter annihilation line by Blackhole of an immediate mass [7][8]. The enhancement factor of the extragalactic neutralino dark matter is one order and three order for $10^2$ M$_\odot$ Blackhole ( scenario A) and $10^5$ M$_\odot$ Blackhole ( scenario B), respectively [8]. Our detector observed 14 and 1440 photons for the 100-GeV extragalactic neutralino dark matter annihilation line that existed at z = 0 by using observations for a period of 3 yr. for scenario A and B, respectively. When we assume same enhancement factor for galactic neutralino dark matter, we observe 300 photons of 100-GeV galactic neutralino dark matter annihilation line for scenario B.

Next, we consider the line sensitivity of the detector. The background for a line is the diffuse gamma ray background. For extragalactic neutralino dark matter annihilation line, statistics of the background for a small energy bin width $\Delta E$ is proportional to $B$ (ph/s/cm²/str) $\times$ $A_{eff}$ $\times$ $T \times \Delta E \times \Omega$. The statistics of the source is $S$ (ph/s/cm²) $\times A_{eff} \times T$. Here, $A_{eff}$ is the effective area; $T$, the observation time; $\Omega$, the field of view; $\Delta E$, the energy resolution. Therefore, the sensitivity $S$ is proportional to $\sqrt{\Delta E / A_{eff} \Omega}$ . We call this value as the relative line sensitivity. For the galactic neutralino dark matter annihilation line, the statistics of the background for a small bin width of $\Delta E$ is proportional to $B$ (ph/s/cm²) $\times A_{eff} \times T \times \Delta E$. The statistics of the source is $S$ (ph/s/cm²) $\times A_{eff} \times T$. $T$ for the Galactic center for an all-sky observation is proportional to $\Omega$. Therefore, $S$ is proportional to $\sqrt{\Delta E / A_{eff} \Omega}$ . We show the energy-dependent relative line sensitivity in figure 10. We obtain a line sensitivity that is 2-3 times better than that of AMS and almost the same as that of GLAST at 10–100 GeV. A COSMOS project shows the three-dimensional distribution of dark matter by calculation of the lens effect of galaxy [9]. However, the COSMOS project dose not show the species of dark matter. As an advantage of high energy resolution, we can measure the red shift of a neutralino dark matter annihilation line. We can obtain the three-dimensional distribution of neutralino dark matter.

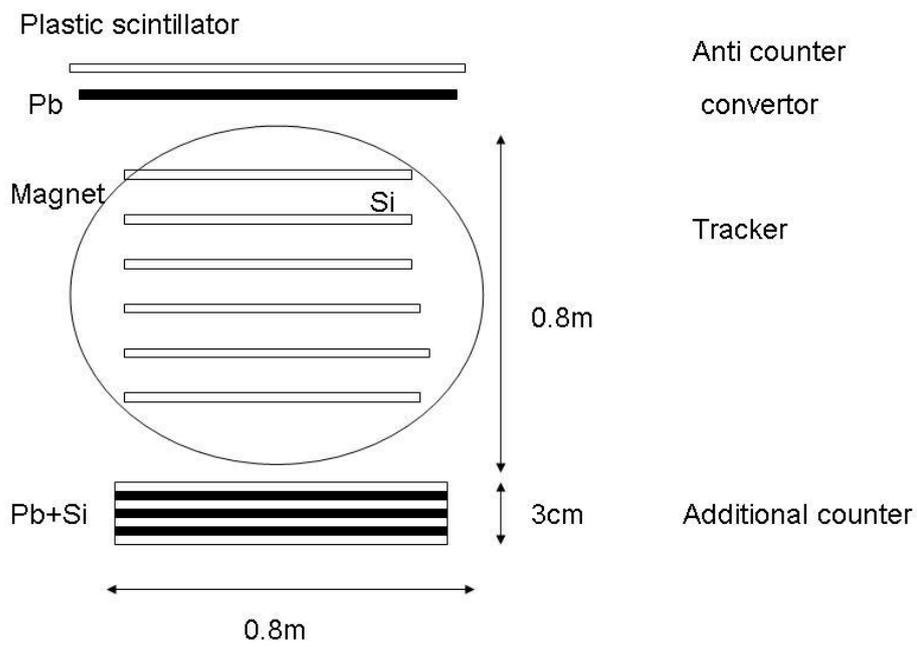

Figure 1: Illustration of GeV gamma ray detector.

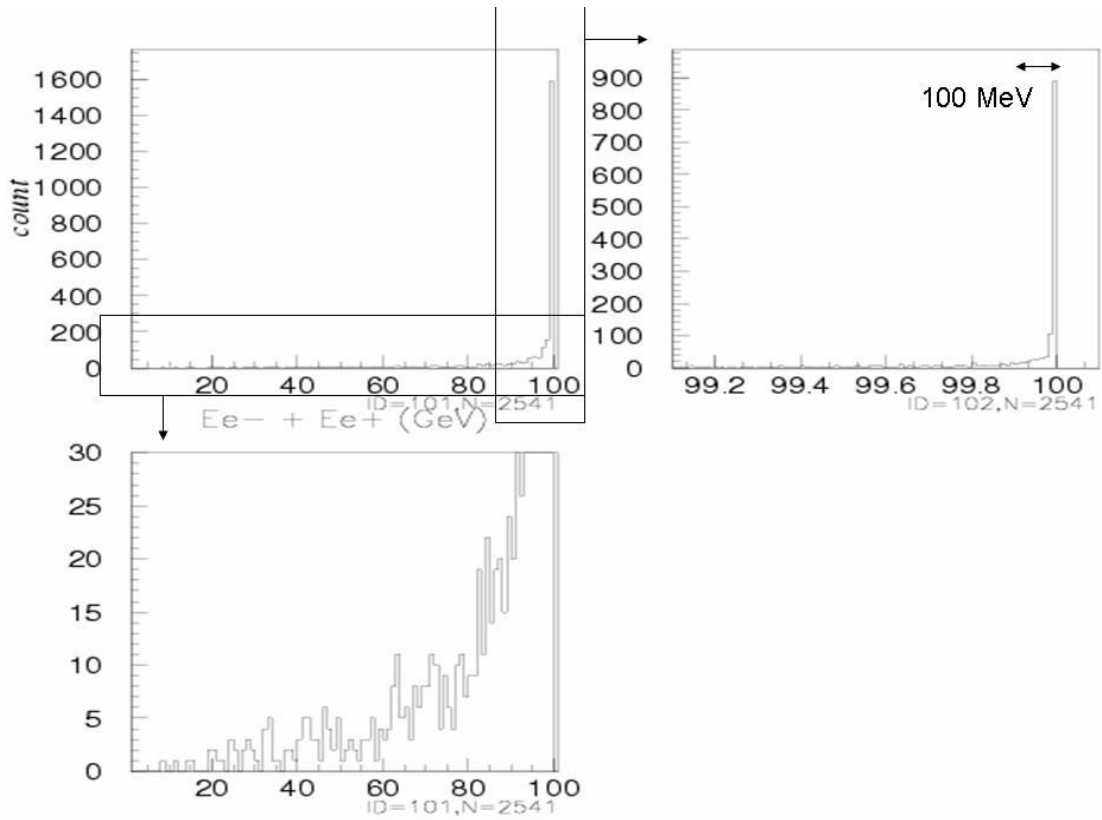

Figure 2: Histogram of the reconstructed energy of an electron-positron pair created in convertor (magnet and cryostat) for a 100-GeV gamma ray; 2541 pairs are created for 30000 gamma rays. Top left:The energy range from 1 to 101 GeV is shown. top right:The energy range from 99.1 to 100.1 GeV is shown. Bottom:The energy range from 1 to 101 GeV is shown and the maximum count in the histogram is set at 30.

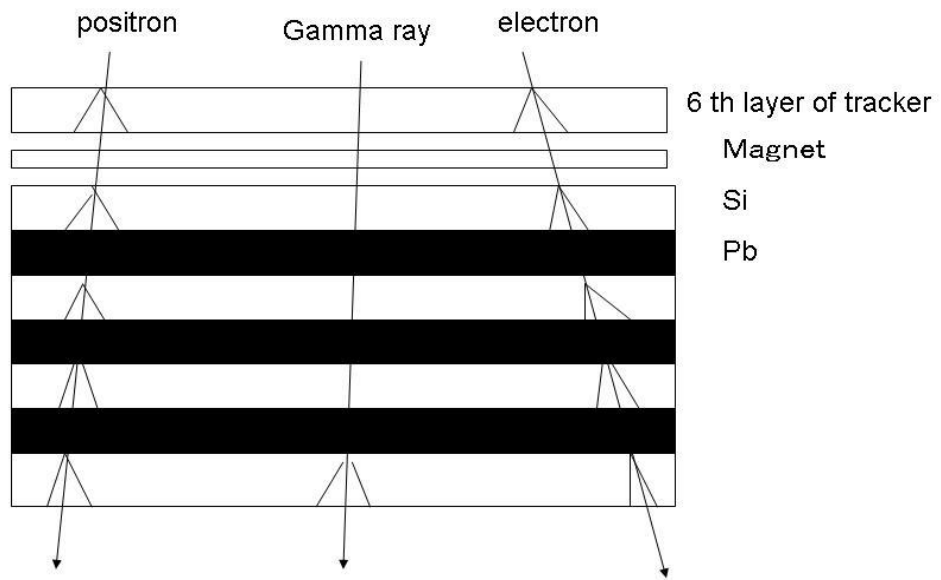

Figure 3: Illustration of an additional counter for the detection of bremsstrahlung gamma rays.

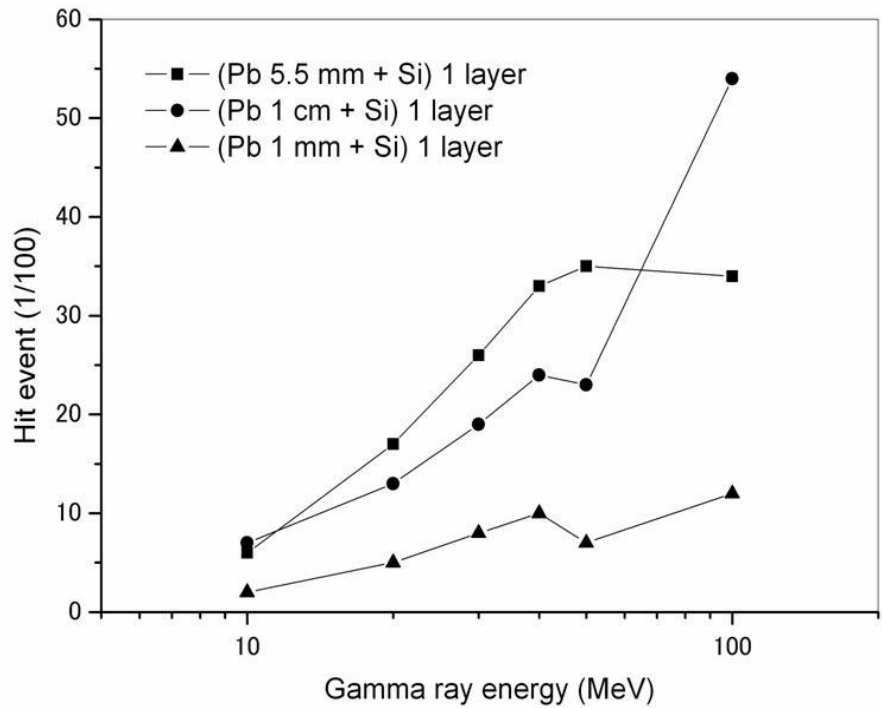

Figure 4: Relation between the number of hits in a Si strip and the energy of gamma rays. 1 layer of a Si strip and a lead is prepared. Three kinds of a thicknesses for the lead—1 mm (0.2 $X_0$), 5.5 mm (1 $X_0$), 1 cm (5 $X_0$) is set. 100 gamma rays as bremsstrahlung is input.

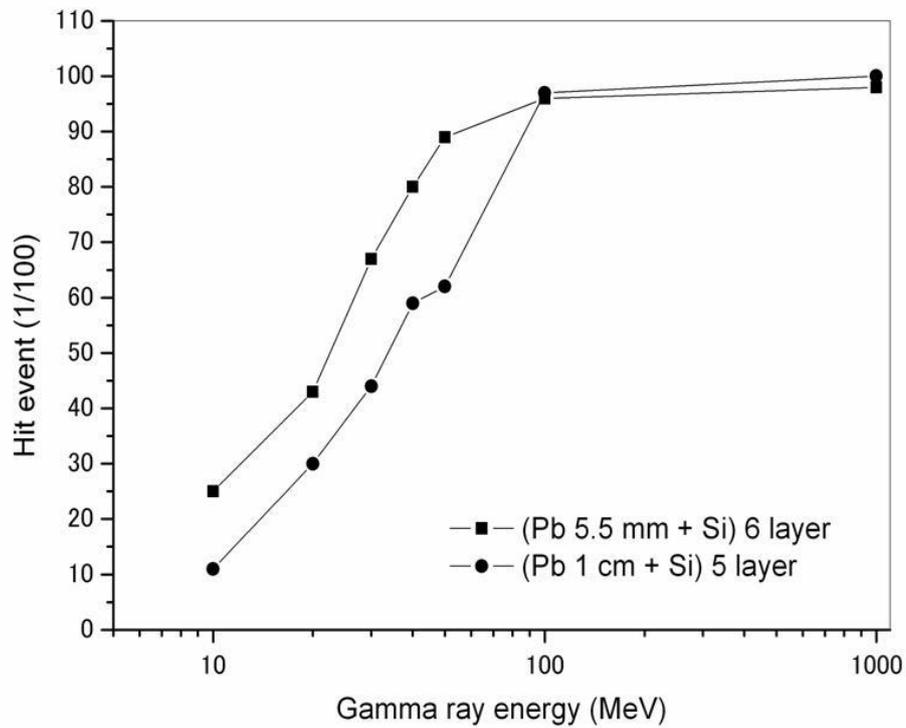

Figure 5: Relation between the number of hits in a Si strip and the energy of gamma ray. Two kinds of counter—six layers of a Si strip and a lead of 5.5 mm and five layers of a Si strip and a lead of 1 cm are prepared. 100 gamma rays as bremsstrahlung are input.

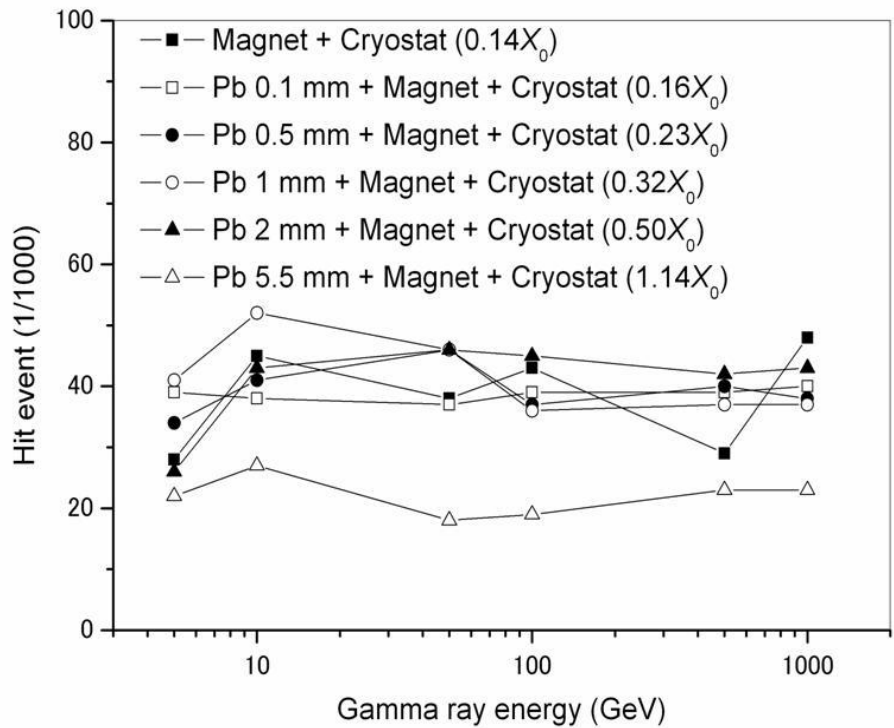

Figure 6 : Relation between the number of selected events and the energy of gamma ray for eight kinds of convertors. The selected condition is that the energy of bremsstrahlung gamma rays is less than 100 MeV, the number of particles is two, and the energy of the particles is above 1 GeV.

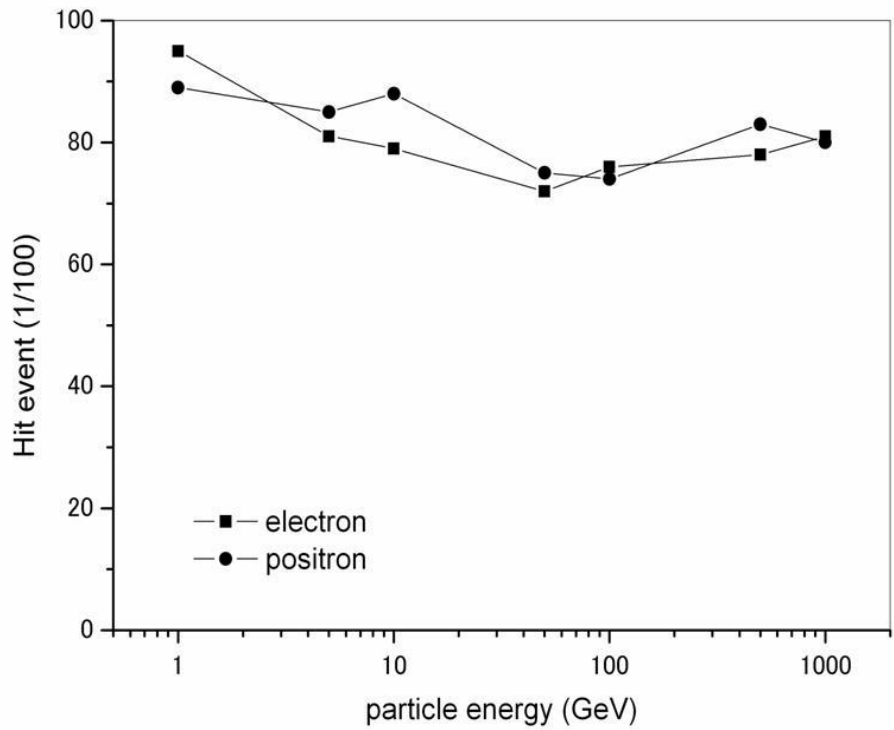

Figure 7:Relation between the number of selected events and the energy of a particle, electron and positron in a Si tracker. The selected condition is that energy of the bremsstrahlung gamma rays is less than 100 MeV and the number of particles is 1.

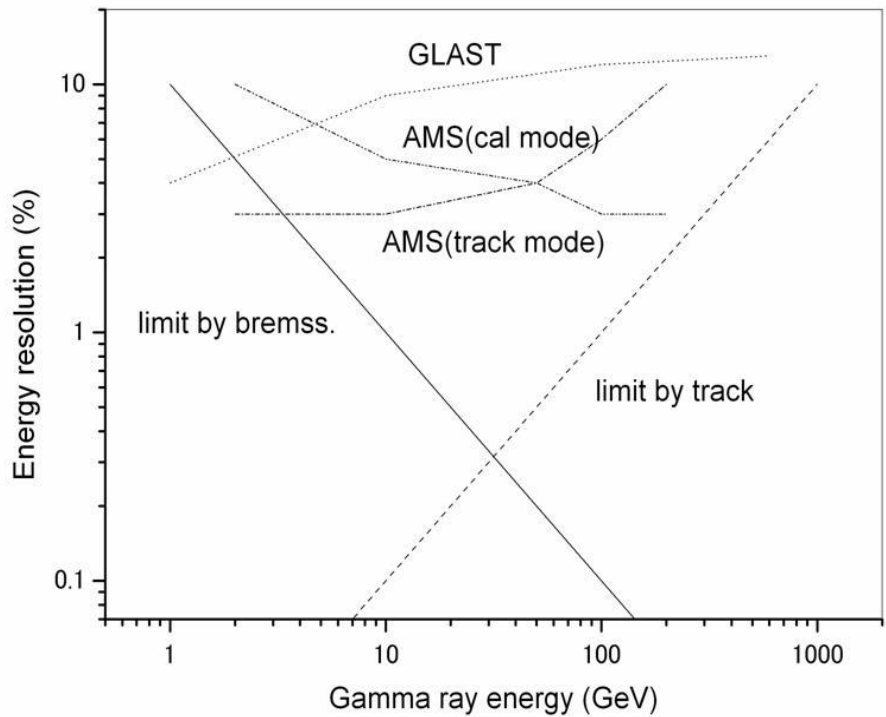

Figure 8: Relation between the energy resolution and the energy of gamma rays. Our detector has an energy resolution allowed by two limits—the accuracy of the energy loss due to bremsstrahlung and the track accuracy.

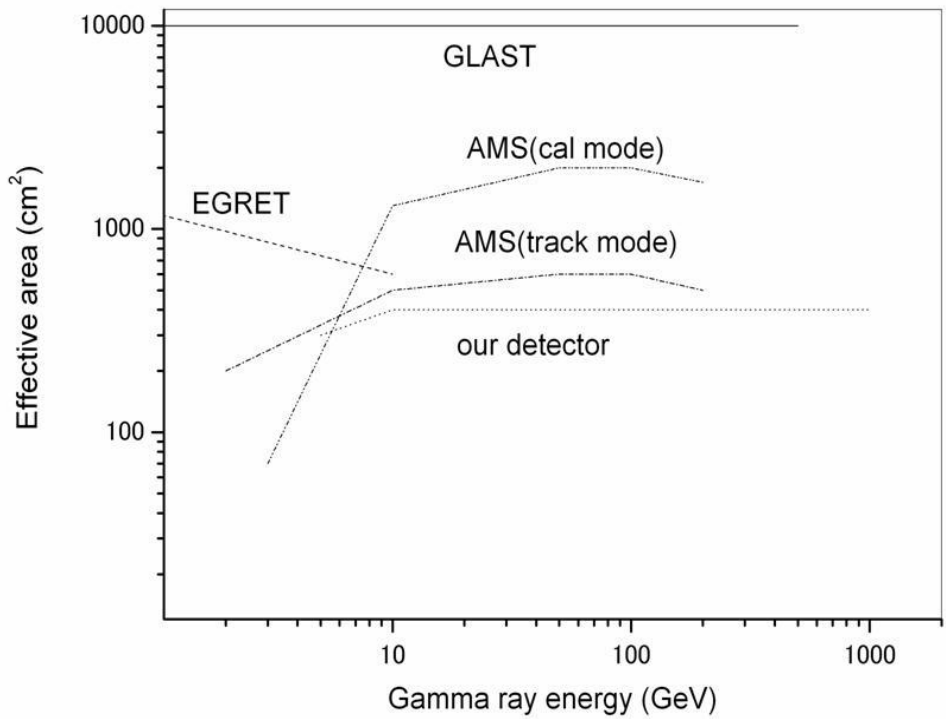

Figure 9: Relation between an effective area and energy of gamma rays.

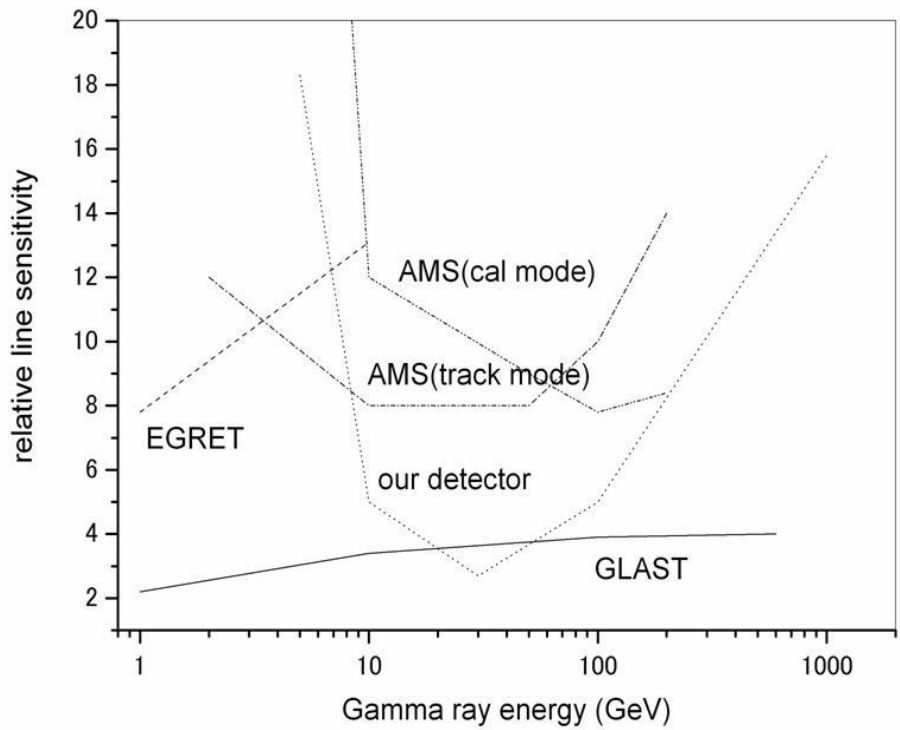

Figure 10:Relation between a relative line sensitivity and energy of gamma rays. The relative line sensitivity is given as $\sqrt{\Delta E / A_{eff}\Omega}$ .